\documentclass{article}

\usepackage{arxiv}

\usepackage{graphicx}
\usepackage{xcolor}
\usepackage[utf8]{inputenc}
\usepackage{rotating}
\usepackage{graphicx}
\usepackage{color}
\usepackage{url}
\usepackage{algorithmic}
\usepackage[boxed]{algorithm}
\usepackage{array}
\usepackage{amsmath}
\usepackage{amssymb}
\usepackage{balance}

\definecolor{mycolor}{RGB}{0,104,152}

\def\CC#1{\textcolor{blue}{\COMMENT #1}}

\def\CC#1{{\textcolor{cyan}{\COMMENT #1}}}

\newcommand{\xLastovka}{{xLastovka}}
\newcommand{\lssOrel}{{lssOrel}}

\newcommand{\genlegendre}[4]{%
  \genfrac{(}{)}{}{#1}{#3}{#4}%
  \if\relax\detokenize{#2}\relax\else_{\!#2}\fi
}

\usepackage{setspace} 

\begin{document}

\title{Computational Search of Long Skew-symmetric Binary Sequences with High Merit Factors}

\renewcommand{\headeright}{}
\renewcommand{\shorttitle}{Computational Search of Long Skew-symmetric Binary Sequences with High Merit Factors}

\newcommand{\orcidlink}[1]{$^{#1}$}


\author{Janez~Brest\,\orcidlink{0000-0001-5864-3533}, 
        and~Borko~Bo\v{s}kovi\'{c}\,\orcidlink{0000-0002-7595-2845}
\thanks{The authors acknowledge the financial support from the Slovenian Research Agency (research core funding No.~P2-0041 -- Computer Systems, Methodologies, and Intelligent Services).}
\thanks{J. Brest, 
and B. Bo\v{s}kovi\'{c} are with the Institute of Computer Science, Faculty of Electrical Engineering and Computer Science, University of Maribor, Koro\v{s}ka c. 46, 2000 Maribor, Slovenia  (e-mail: \{janez.brest,\,borko.boskovic\}@um.si).}}

\markboth{Journal of \LaTeX\ Class Files, Vol. 14, No. 8, August 2015}
{Shell \MakeLowercase{\textit{et al.}}: Bare Demo of IEEEtran.cls for IEEE Journals}
\maketitle


\begin{abstract}
In this paper, we present a computational search for best-known merit factors of longer binary sequences with an odd length. 
Finding low autocorrelation binary sequences with optimal or suboptimal merit factors is 
a
very difficult optimization problem. 
An improved version of the heuristic algorithm is presented and tackled to search for aperiodic binary sequences with good autocorrelation properties.
High-performance computations with the execution of our stochastic algorithm
to search skew-symmetric binary sequences with high merit factors. 
After experimental work, as results, we present new binary sequences with odd lengths between 201 and 303 that are skew-symmetric and have the merit factor $F$ greater than 8.5. 
Moreover, an example of a binary sequence having $F > 8$ has been found for all odd lengths between 201 and 303.
The longest binary sequence with $F>9$ found to date is of length 255.
\end{abstract}

\keywords{Golay's merit factor \and binary sequences \and aperiodic autocorrelation sidelobes \and skew symmetry}

\maketitle

\section{Introduction}
\label{Sec:Intro}
%
%
%

Binary sequences with low autocorrelation function properties are important in many areas, such as communication engineering~\cite{2017TutorialSeq,zhao2017unified,ukil2015asymptotic,zeng2020optimal} and in
statistical mechanics~\cite{Lib-OPUS-labs-1996-JPhysA-Mertens-BB_solutions,Lib-OPUS-labs-1987-JourPhys-Bernasconi,2015Leukhin,tomassini2021complex}. Also in mathematics, this problem (see {\it Littlewood polynomial}) has attracted sustained interest~\cite{GUNTHER2017340,JEDWAB2013882,gunther2018flat}.
Long binary sequences are essential for various applications of the coded exposure process~\cite{jeon2013fluttering,jeon2017generating}. 

Finding Low Autocorrelation Binary Sequences (LABS) with optimal/good merit factors or peak sidelobe level is a challenging optimization problem. There are two types of binary sequences: periodic and aperiodic. In this paper, we are dealing with aperiodic ones.

A binary sequence $S = (s_0,s_1, \ldots, s_{L-1})$ has all entries either $+1$ or $-1$. Here, $L$ denotes the sequence length. 
The aperiodic autocorrelation of $S$ at shift $k$ is defined as:
\begin{multline}\label{eq:ck}
C_k(S) = \sum_{i=0}^{L-k-1}s_{i}s_{i+k}, \\
{~\textrm{for}~} k = -(L-1), \dots, -1, 0, 1, \dots, L - 1,
\end{multline}
and the Integrated Sidelobe Level (ISL) metric of $S$ is:
\begin{equation}
\textrm{ISL}(S) =  \sum_{k=1}^{L-1}|C_{k}(S)|^2.
\label{eq:labs:energy}
\end{equation}
Note that $\textrm{ISL}(S)$ is defined as the sum of the squares of all off-peak autocorrelations (i.e., $k \ne 0)$.

The LABS problem involves assigning values 
to the $s_i$ that minimize $\textrm{ISL}(S)$ or maximize the {\em merit factor $F(S)$} 
\cite{Lib-OPUS-labs-1977-IEEE_IT-Golay}:
\begin{equation}
F(S) =  \frac{L^2}{2 \cdot \textrm{ISL}(S)}.
\label{eq_meritFactor}
\end{equation}
The merit factor $F(S)$, shortly $F$ in the remaining of the paper, is a measure of the quality of the sequence in terms of engineering applications~\cite{Borwein:2008merit}.

The {\em skew-symmetric sequences} have odd length with $L=2n-1$ and satisfy:
\begin{equation}
s_{n+i} = (-1)^{i}{s_{n-i}}, ~ i=1,2,\ldots, n-1.
\label{eq:skewSymmetric}
\end{equation}
which implies that $C_k(S)=0$ for all odd $k$.
The restriction of the problem to skew-symmetric sequences reduces the sequence's effective length from $L$ up to approximately $L/2$. It means that the dimension of the problem and the search space are reduced. The search space is reduced from $2^L$ to approximately $2^{(L/2)}$~\cite{Lib-OPUS-labs-1996-JPhysA-Mertens-BB_solutions}. Note that the optimal skew-symmetric solutions might not be optimal for the whole search space.

Besides the merit factor, another metric for the LABS problem is the Peak Sidelobe Level, $\textrm{PSL}(S) = \max_{k = 1}^{L-1}  | C_k(S) |$~\cite{jedwab2006peak}. 
Most of the time, a sequence with the optimal PSL has a merit factor which is much lower than the optimal merit factor, and vice versa.  
In this paper, our key focus is to search
for long aperiodic binary sequences with high merit factors. 
A reader interested in optimization of the PSL values is referred to works~\cite{dimitrov2020generation,brest2021low,dimitrov2020aperiodic,dimitrov2021hybrid,chen2021computationally}. 

One of the main challenges when solving the LABS problem using the incomplete search is how to 
implement a calculation of energy in Eq.\,(\ref{eq:labs:energy}) efficiently. 
Researchers developed an efficient implementation of the energy calculation~\cite{Lib-OPUS-labs-2009-ASC-Gallardo-memetic,Lib-OPUS-labs-2008-CP-Halim-tabu,BoskovicLABS,lin2019efficient,dimitrov2020efficient,dimitrov2020generation}. Note that a similar efficient calculation can be also applied to finding a skew-symmetric solution of the odd length problem instances~\cite{BoskovicLABS,dimitrov2021skew}.

For the time being, aperiodic binary sequences with currently known best merit factors for lengths from 191 up to 225 are published in~\cite{Brest2018heuristic}. All these sequences are skew-symmetric with $8.6394 < F < 9.5851$.
For lengths longer than 225 up to 301, there are known sequences for some lengths only and all of them have $F<8$ (see collection~\cite{BoskovicLABS}).
Searching binary sequences, general or skew-symmetric, with a high merit factor, higher than 8 for a length longer than 230 is a challenging optimization problem.

Nowadays, parallel computation can be applied to tackle hard optimization problems. The power of multiple computers, which are not necessarily placed in the same location but can be distributed, is combined to solve multiple real-world problems.
Grid computing is used in literature to make computations for finding (binary) sequences~\cite{packebusch2016low,katz2020crosscorrelation,BoskovicLABS,Brest2018heuristic}.

In this paper, we use 
an improved version of the \xLastovka~\cite{Brest2018heuristic} stochastic algorithm for searching 
skew-symmetric binary sequences. 
In particular, we investigate through extensive experimental  runs the influence of dimensionality of binary sequences for odd lengths $225 < L \le 303$.
At the end of this experimental work, we are 
able to find a number of binary sequences that have merit factors higher than 8.
The main contributions in this paper can be summarized as follows:
\begin{itemize}
    \item The improved version of the algorithm has found skew-symmetric binary sequences with the same or better merit factor than previous algorithms for lengths between 201 and 225.
    \item For all lengths between 227 and 303, including, we have been able to find skew-symmetric binary sequences with $F>8$.
    \item Examples are now known of binary sequences with $201 \le L \le 281$ and $L = 285$ having merit factors greater than 8.5.
    \item The longest skew-symmetric binary sequence with $F>9$ found to date is of length $L=255$.
\end{itemize}

The rest of our paper is organized as follows.
The background is given in Section~\ref{sec:Background}, where related work is also presented.  
In Section~\ref{sec:Algorithm} an algorithm for solving a LABS problem in the sense of searching skew-symmetric binary sequences with high merit factor values is presented.
Section~\ref{sec:Results} is the main part of the paper, where 
experimental results are conducted, and a brief discussion is given. 
Finally, the paper ends with a conclusion
and future work in Section~\ref{sec:Conclusion}.

\section{Background}
\label{sec:Background}

Theoretical considerations from Golay in 1982~\cite{Lib-OPUS-labs-1982-IEEE_IT-Golay} give an upper bound on $F$ of approximately 12.3248 as $L \rightarrow \infty$. However, Golay 
does not prove that 12.3248 is an upper bound on the asymptotic merit factor, because it relies on an unproven heuristic argument.

Owing to the practical importance and widespread applications of sequences with good autocorrelation properties, 
in particular with low peak sidelobe level values 
or large merit factor values, a lot of effort
has been devoted to identifying these sequences either by analytical construction methods or computational approaches in the literature~\cite{song2015optimization,zhao2017unified,tarnu2022maximal}.  

The {\em construction method} is set by so called appended rotated Legendre sequences with an asymptotic merit factor of  6.342061...~\cite{JEDWAB2013127,JEDWAB2013882}.
On the other hand,~\cite{Baden:2011} used the modified Jacobi sequences together with the steep descent algorithm, and got an approximate asymptotic merit factor of 6.4382. 
The gap toward Golay's upper bound, i.e., 12.3248, still remains huge.
Notice, that the study of the merit factor is fundamentally concerned with an asymptotic behavior, and not the identification of a particular sequence with a large merit factor.
Nevertheless, J. Jedwab in the survey~\cite{jedwab2004survey} gave 
 a personal selection of challenges concerning the Merit Factor problem, arranged in order of increasing significance.  The first challenge is as follows:
``{\it Find a binary sequence $S$ of length $L > 13$ for which $F \ge 10$.}''
Interestingly, in 2005, R. Ferguson and J. Knauer~\cite{ferguson2005optimization} suspected that in lengths of perhaps 250 for skew-symmetric sequences that merit factors $F>10$ will regularly start to appear.  To find a general or skew-symmetric binary sequence with $F \ge 10$ still remains open. 

The search space of the LABS problem is of size $2^L$.
To locate good (optimal) solutions, two approaches exist: {\em complete} and {\em incomplete} search. 
 The complete, or exact search, 
is able to find the optimal sequence, but it is unlikely to scale up to large sequences.
The incomplete, or stochastic search, can obtain a result that may be optimal or close to optimal, i.e., it does not guarantee  optimality.

Currently, the {\em optimal solutions} for binary sequences of even and odd lengths are known for $L \le 66$, calculated by T.~Packebusch and   S.~Mertens in 2016~\cite{packebusch2016low}.  
Interestingly, it took 20 years to prove optimality for six sequences with $61 \le L \le 66$.
Optimal solutions were proved by using the branch-and-bound algorithm.

Following the theoretical minimum energy level analysis, a new
asymptotic merit factor value of 10.23 was {\it estimated} by Ukil~\cite{ukil2015asymptotic} in 2015 based on sequences of
length 4 to 60, found by the exhaustive search. 

The {\em optimal skew-symmetric solutions} are known for 
$L \le 119$~\cite{packebusch2016low}. The previous record was $N \le 89$~\cite{Lib-OPUS-improved-branch-and-bound-for-low-autocorrelation-binary-sequences-2013-Prestwich}.

On the other hand, 
heuristic algorithms were introduced for solving many real-world problems. 
A heuristic algorithm can solve small instances easily and performs reasonably when tackling larger instances by quickly finding solutions that are close to optimal. 
The optimality of the solution is not guaranteed as in, for example, the exhaustive search.

Different techniques have been utilized to tackle the LABS problem, 
such as enumeration~\cite{Lib-OPUS-labs-1992-Optimization-deGroot}, 
evolution strategy~\cite{Lib-OPUS-labs-1992-Optimization-deGroot}, 
genetic algorithm~\cite{1993-NBDT-Reinholz}, 
local search algorithm~\cite{farnane2018local}, 
branch and bound~\cite{Lib-OPUS-labs-1996-JPhysA-Mertens-BB_solutions,packebusch2016low}, 
evolutionary algorithm with a suitable mutation operator~\cite{Lib-OPUS-labs-1998-IEEE_EC-Militzer}, 
tabu search~\cite{Lib-OPUS-labs-2008-CP-Halim-tabu},
directed stochastic algorithm~\cite{Borwein:2008merit}, 
evolutionary algorithm~\cite{Deng1999new},
memetic algorithm combined with tabu search~\cite{Lib-OPUS-labs-2009-ASC-Gallardo-memetic}, 
and self-avoiding walk technique~\cite{BoskovicLABS,Brest2018heuristic}.

The memetic agent-based paradigm~\cite{zurek2017toward}, which combines evolutionary computation and local search techniques using parallel GPU implementation, is one of the promising meta-heuristics for solving a LABS problem. 

\begin{figure}[ht]
\centering
\includegraphics[width=\linewidth]{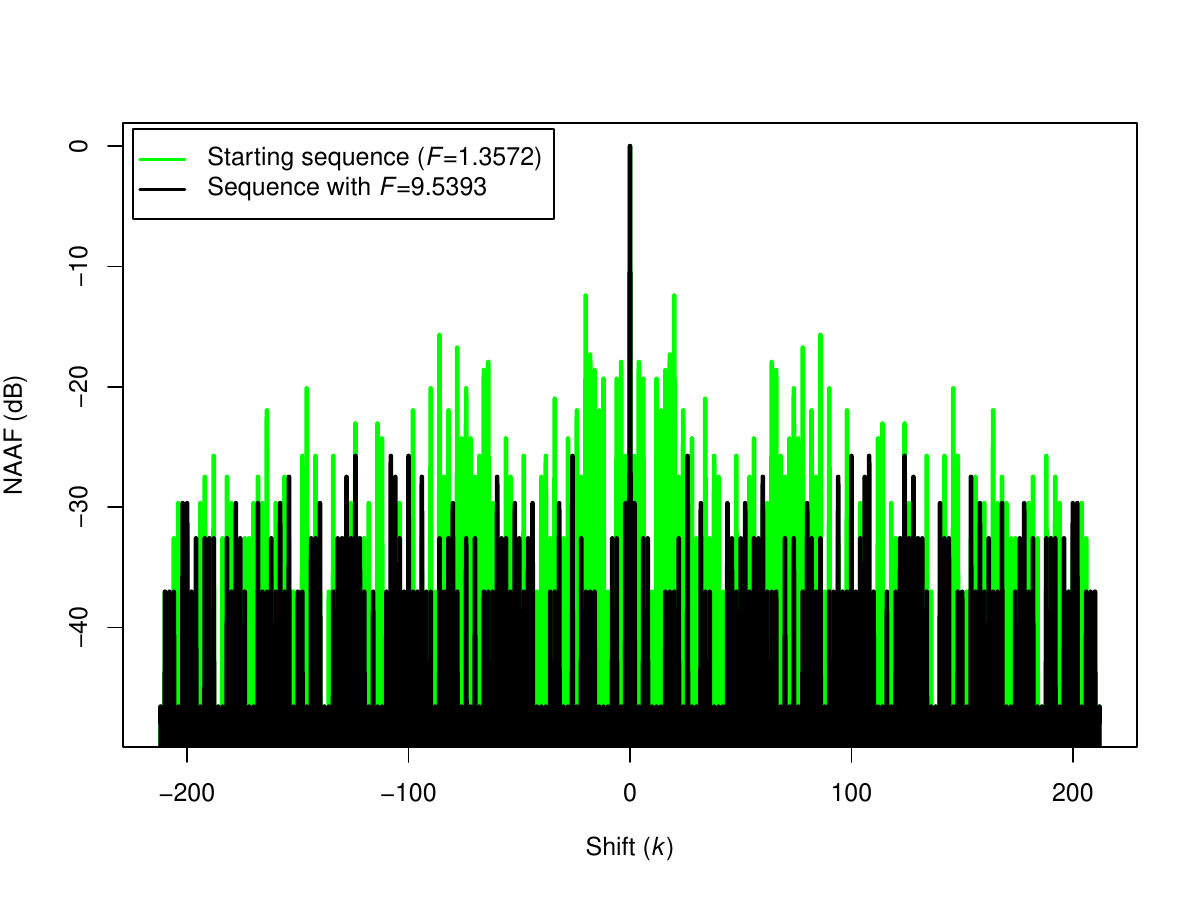}
\caption{A binary sequence of length $L=213$ with $F$ = 9.5393 and a corresponding starting binary sequence.}
\label{fig:L213:NAAF}
\end{figure}

Figure~\ref{fig:L213:NAAF} shows the normalized aperiodic autocorrelation function (NAAF) in dB, i.e., 20log$_{10}\frac{C_k(S)}{L}$, of two binary sequences of length 213. One is randomly generated with $F$ = 1.3572, and another has $F$ = 9.5393, which is currently the best-known merit factor for sequences with a length over 200.  The NPSL values, i.e., 20log$_{10}\frac{\footnotesize{\textrm{PSL}}(S)}{L}$,  of the randomly generated sequence and the sequence with $F$ = 9.5393 are $-12.42$ dB, and $-25.74$~dB, respectively, and the optimized sequence has the NPSL value, which is more than 13 dB lower than that of the starting sequence. 

As we already said, a sequence with the optimal PSL usually has a merit factor that is much lower than the optimal merit factor, and vice versa. For example, the sequence with $F$ = 9.5393 (Fig.~\ref{fig:L213:NAAF}) has $\textrm{PSL}=11$, while a sequence with the ``good'' $\textrm{PSL}$ value of 9 has the merit factor $F=4.8386$, as reported in~\cite{brest2021low}. 

Notice, that the \lssOrel~\cite{Brest2018heuristic} algorithm belongs to a group of the deep first search algorithms (after flipping each $s_i$ it continues with the best sequence from $s_i$ only), and it applies restarts (randomly initialized new starting sequence) after a predefined number of deep first search steps.

M.~Dimitrov et al.~\cite{dimitrov2021skew}, recently proposed an algorithm with a similar technique as used by \lssOrel, with the difference that in~\cite{dimitrov2021skew}, the algorithm starts a new search with the sequence of flipped $s_i$ immediately, when $s_i$ improves the merit factor. This algorithm applies small perturbations of flipping a few bits to continue the search after all flipped $s_i$ did not improve the merit factor. It uses an efficient calculation of merit factors by the storing pre-calculated values of $C_k$ in a one-dimensional array.
This algorithm was able to perform a computational search on skew-symmetric sequences with lengths up to $10^5+1$ and obtained merit factors $F \approx 5$. For sequences with  $200<L<300$, there are no reported sequences with merit factors greater than 7; the sequence with $F=6.5319$ is reported for $L=449$~\cite{dimitrov2021skew}.
Therefore, finding examples of aperiodic binary sequences with $200 \le L \le 300$ that have merit factors greater than 8 or even greater than 9 is very challenging.

\section{Algorithm impxLast}
\label{sec:Algorithm}

In this section, we give an overview of the algorithm that was used in experimental work in this paper. We present the algorithm's features that are necessary for solving the LABS problem effectively and efficiently. 
 In the experimental work, we used an improved version of the \xLastovka~\cite{Brest2018heuristic} algorithm.  We will call the improved version of the algorithm, impxLast. There are differences in the implementation of both algorithms. The \xLastovka\ algorithm uses two hash tables, one for storing visited sequences, i.e., we say that a sequence is visited when its merit factor is calculated (Eq.~(\ref{eq_meritFactor}) in general), and another for storing visited one-bit flipped neighborhood sequences. The main reason for using the second hash table lies in avoiding the already visited sequences and unnecessary calculating of their merit factors more times, since the LABS problem with skew-symmetric search space is very likely making cycles (i.e. repeating of already visiting sequences) when searching neighborhood sequences.
 
 In the implementation of our improved algorithm,  we omit the second hash table and, consequently, the impxLast algorithm requires more calculations of the merit factor values in comparison to the original algorithm, but on the other hand, a new algorithm can therefore be slightly faster in the sense if we are focusing on the speed of both algorithms. The speed is defined as the number of function evaluations per second. A one-bit flip operation is counted as one function evaluation. 
 
 The mentioned difference between both algorithms plays a key role in making the new algorithm more successful in finding longer binary sequences with larger merit factor values. A greater amount of frequent accesses to the hash table can also reduce the performance of the algorithm. 
 In our algorithm, we store one-bit flipped neighborhood sequences into a priority queue based on their merit factors, i.e., the sequences with promising merit factors are inserted into a priority queue. Since the priority queue has a fixed size, a sequence with the worst merit factor is removed from the priority queue when a new sequence with higher merit is inserted into the priority queue. The same mechanism of storing one-bit flipped neighborhood sequences is also applied in the xLastovka algorithm.

A pseudocode of the impxLast algorithm is presented in Algorithm~\ref{alg_MFlastovkaDva}. It can be viewed as the best first search algorithm. 
In the main loop in our algorithm, each $s_i$ is flipped, and its merit factor is calculated, then this new sequence is inserted into a priority queue, and, finally, it is compared to the sequence with the currently best merit factor. If necessary, the algorithm saves the new best sequence and its merit factor. After each $s_i$ is flipped, the algorithm removes the sequence with the highest merit factor from the priority queue and continues the search process using this sequence as a new starting sequence. 
 At the end of the search process, the algorithm outputs saved the best skew-symmetric binary sequence and its merit factor value.
 
 A fast calculation merit factor of a sequence with one-bit flipped is used in Step~\ref{alg:fastCalcMF} in Alg.~\ref{alg_MFlastovkaDva}. This mechanism was proposed in~\cite{Lib-OPUS-labs-2009-ASC-Gallardo-memetic} and it uses two-dimensional structure for the efficient calculation of a merit factor when one bit is being changed.  The same mechanism is also applied in~\cite{Brest2018heuristic}. This mechanism is very suitable for not very long binary sequences, i.e. $L$ up to a few thousand, since its space complexity is $O(L^2)$. Note that Dimitrov et al.~\cite{dimitrov2021skew} also proposed an efficient mechanism for one-bit flip of a skew-symmetric binary sequence, which has space and time complexity $O(L)$ and it is suitable for searching very long binary sequences. We refer to~\cite{dimitrov2021hybrid,brest2021low} for a couple of recent developments in searching of  binary sequences with low PSL.
 In Step~\ref{alg:step:forLoop}, $(L+1)/2$ flips are performed for each bit of a skew-symmetric binary sequence. The outer while loop that starts in Step~\ref{alg:step:while} is executed while stopping criteria are not met. In the experimental work, we used a time of four days as the stopping condition in the impxLast algorithm.


\renewcommand{\algorithmiccomment}[1]{#1}

\begin{algorithm}[tb]
\caption{Best first search algorithm (impxLast) with a priority queue.}\label{alg_MFlastovkaDva}
\begin{algorithmic}[1]
\begin{small}
\REQUIRE \COMMENT {$L$ ... length of sequence}\medskip\\
\ENSURE $S_{best}$  ... best sequence found during optimization search  
\ENSURE $F_{best}$  ... merit factor value of $S_{best}$\smallskip  

\STATE $S \gets $  initialize a starting skew-symmetric sequence of length $L$ 
\STATE $F \gets $ calculate merit factor using Eq.~(\ref{eq_meritFactor})
\STATE Insert $S$ into hash table HT 
\STATE $S_{best} \gets S$;  $F_{best} \gets F$
\WHILE{stopping criteria are not met} \label{alg:step:while}
   \item[] \CC{** search neighborhood **}
   \FOR{ each $i \in (L+1)/2$} \label{alg:step:forLoop}
      \STATE $S_f \gets $ flip $s_i$ in $S$
      
      \IF{$S_f \in $ HT} 
         \STATE {\bf continue} \CC{** skip if sequence $S_f$ was already visited **}
      \ENDIF
      \STATE $F_f \gets $ fast calculate merit factor of $S_f$ (skew-symmetric) \label{alg:fastCalcMF}
      \STATE Insert $S_f$ into a priority queue ordered by merit factor
      \IF{$F_n > F_{best}$} 
         \item[] \CC{** save the best sequence and its merit factor **}
         \STATE $S_{best} \gets S_f$; 
 $F_{best} \gets F_f$
      \ENDIF
   \ENDFOR
   \item [] \CC{** best first search **} 
   \STATE $S \gets $ remove an item from the front of the queue
\ENDWHILE \medskip \\

\end{small}
\end{algorithmic}
\end{algorithm}

\section{Results and Discussion}
\label{sec:Results}

\begin{figure*}[htb]
\centering
\includegraphics[width=0.99\linewidth]{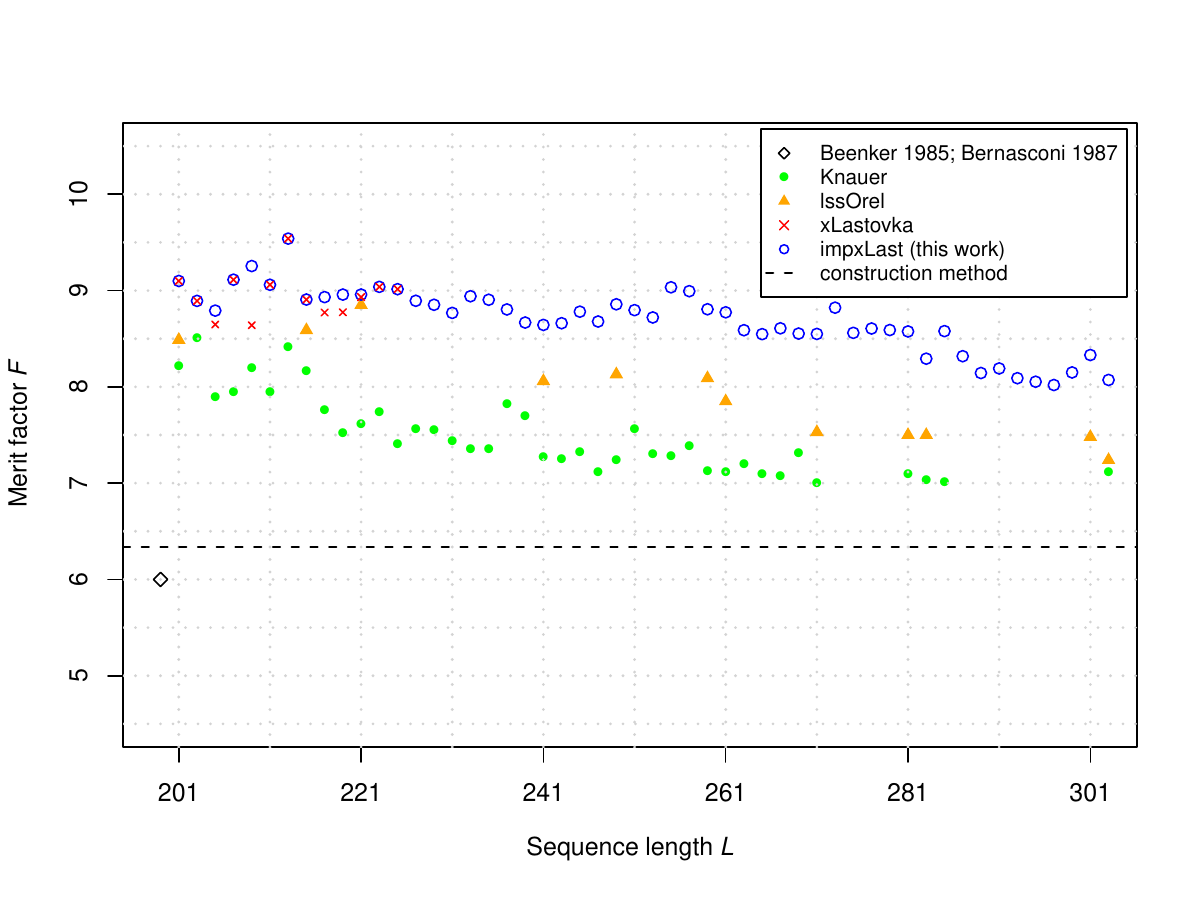}
\caption{Merit factor values for skew-symmetric binary sequences with odd length from 201 up to 303. Beenker \& Bernasconi's result is for length 199.}
\label{fig:MF-201-303}
\end{figure*}

In this section, we present our main results with  discussion.
The result for $201 \le L \le 303$ obtained by the impxLast algorithm are presented in Table~\ref{tab:best:L201}, and the merit factors of the obtained binary sequences are plotted in Figure~\ref{fig:MF-201-303}. In Table~\ref{tab:best:L201}, the sequences obtained by the \xLastovka~\cite{Brest2018heuristic} algorithm are marked  with $^\dagger$, and the impxLast algorithm was able to find a sequence with the same merit factor value, too. 

Figure~\ref{fig:MF-201-303} shows that all currently best-known merit factor values for skew-symmetric binary sequences with the odd length between 201 and 303 are greater than 8. 
With a construction method based on rotated Legendre sequences, one can construct a sequence with any arbitrary length (usually these lengths are required to be prime numbers) that has a merit factor of approximately  6.34~\cite{Borwein:2004:merit:6.34,Baden:2011}. We can notice gab of 2 between currently best-known merit factors obtained by the impxLast algorithm and merit factors generated by the construction method on Figure~\ref{fig:MF-201-303}.

The best-known merit factor values in Figure~\ref{fig:MF-201-303} are decreasing for longer sequences.  
A trend that the merit factor $F$ decreases as length $L$ increases can be interpreted in a way that a sequence with a high merit factor is harder to be found by a heuristic algorithm as its length increases. Note that our algorithm used the same amount of execution time for each sequence length. 
We used the SLING infrastructure~\cite{Lib-OPUS-SLING-2020} to perform one hundred runs for each LABS instance, and for some instances {with lengths} around 290, more executions were required to reach $F>8$. Each run was limited to 4 days.

Figure~\ref{fig:MF-201-303} also illustrates how the best-known merit factor values have been changing over the last 35 years. The best-known merit factor was approximately 6 in 1985~
 \cite{Lib-OPUS-labs-1985-Phillips-Beenker,Lib-OPUS-labs-1987-JourPhys-Bernasconi} 
 for skew-symmetric sequences up to $L=199$.
 Knauer's results are dated back to 2004, which is roughly speaking 2 decades after Beenker's results, and our results are approximately 2 decades after Knauer's results. This indicates how hard is the computational search of the LABS sequences with high merit factors.
 
\renewcommand{\arraystretch}{1.30}
\begin{table*}
\centering
\caption{Merit factors and skew-symmetric binary sequences of length 201 up to 303 obtained by the impxLast algorithm. Sign $^\dagger$ indicates the sequence and its merit factor that was also obtained  in~\cite{Brest2018heuristic}.}
\label{tab:best:L201}
\resizebox{\textwidth}{!}{
\begin{small}
\begin{tabular}{clrc}
\hline
$L$ & $F$ & Sequence \hspace{5cm} & PSL \\
\hline
201 & 9.0993$^\dagger$ & 00FC7C04C5DF914630C9E3AF60A741258CB26FB95DECEAD6D4A & 15 \\
203 & 8.8927$^\dagger$ & 0E3C71E0783B8073005132B88CDDBF310553255BB5A56924B6D & 13 \\
205 & 8.7918 & 06333981E3DC3FDAA0FAF0E0106BAA4B414A01D52DD25A993326 & 15 \\
207 & 9.1129$^\dagger$ & 0492402193C9DA4ED4207EED42740EEA5740EC789CB1975471C5 & 19 \\
209 & 9.2544 & 071C7077C4519F3303F82A181F5F5A9A02952B3359BEED76B6DB6 & 15 \\
211 & 9.0600$^\dagger$ & 0E38E0EF88A33E6607F054303EBEB534052A5666B37DDAED6DB6D & 15 \\
213 & 9.5393$^\dagger$ & 0545F75480D9EAD2791B136CB1E25B0C73B1B963C059CA8FF75EFE & 11 \\
215 & 8.9066 & 03E1B0FB2086FA3FE0628366088DDD6635F656AB7AE5D73AD396B5 & 13 \\
217 & 8.9319 & 04A40097BF6B77C98C7BD94E876DC7684F9D16C98D77075178AAE0E & 19 \\
219 & 8.9580 & 03E0F983E0CDF0067745479A682475E79A4042265528CD6B59AD6B5 & 15 \\
221 & 8.9584 & 00037B10994896495AC2928C7E79D9656C8382C1F8E788F98BB172AA & 17 \\
223 & 9.0383$^\dagger$ & 03E3E1F477C04FF98F31B71E403113546923932D9AAC54A24296B6B5 & 13 \\
225 & 9.0144$^\dagger$ & 007030301EFF79C03DC6E7347AC71B6C16F3646DD2AD97545AB2B2B6A & 15 \\
227 & 8.8935 & 1CEE773970A8710C03C05C7AE04616456FA4854B54D125FD21B226EC9 & 15 \\
229 & 8.8523 & 048B2681508906F64BA3F05CF4E0C8CA4F4DEB5210E746B88BFA86308E & 13 \\
231 & 8.7678 & 043FFF7E313D68635219C13C86706E5265CB149970365E0B136A2AAB45 & 15 \\
233 & 8.9409 & 038E3387551EA3F15BC0D5911D4867668FDBB9FCAD1FB5205BFF6932492 & 15 \\
235 & 8.9044 & 0DBB2327E024F9E6AE8AB42CF3C85405CB2CF43FDEFE69AC756A7373B8D & 15 \\
237 & 8.8039 & 03E24D9306D80F80E5C4082B0115D8C9DFBAB028AEDE4A94A9C6B39CE252 & 15 \\
239 & 8.6678 & 1CD29A646F01F337DAEB55425A4204C5747874003EF8A332952E4679F0C9 & 13 \\
241 & 8.6430 & 054AA555344A868922C49C36652531B1B3E3E672D8EC2388680EF3FFE00FE & 15 \\
243 & 8.6608 & 0383FE107E1E00F38E260034CCEFB8BDBAECCC355676DB2D5696A516AB5B5 & 17 \\
245 & 8.7807 & 00502FC5B97D6B64C8863D348038918C9B892A8F3D2688CE707D791ED42BEA & 13 \\
247 & 8.6784 & 11810DC8CA02F4E19042E9658F71DE2F68922D861EF45196C2F57CDC8D1591 & 15 \\
249 & 8.8573 & 000FF810253878C6360C380BD9EFDBB1B11D459D0A92CA726C9693E2BA954AA & 17 \\
251 & 8.7966 & 03B8E01DC783BE30F1B898A23FC4560356044AB77D9DB92D36BB5A48956DBB5 & 17 \\
253 & 8.7206 & 0780041F0DE2791BFC10E4F2616C8C4828EC8C7A634E4BAD51B9625CB5AEAA96 & 13 \\
255 & 9.0338 & 00FE00B692427B7F924094B3398FB9EE6E9BAD9B33C1D471AA3A7471E3D56AD5 & 19 \\
257 & 8.9936 & 01B2CBE1F07DB2728B67E44A8C0730005EAAB36AC80EE567083631D6B5A50C31A & 19 \\
259 & 8.8056 & 1C3C6317188205030EBF1D99E819EE54BC06E995E99892BED350575D921364B49 & 17 \\
261 & 8.7740 & 03F9F0E0184FB6387379B5C1522544E757F64EFE23FADF1973692714E9AA4B5952 & 15 \\
263 & 8.5882 & 0C71C638FC545E3B2D6F22B4AA667F91FA91AA667FC3F72E0F3B68404ADB64924D & 15 \\
265 & 8.5473 & 0000489366D6AC48B59B2D207972A7E5432FE5603796A3C319F08EC07C67388EAAA & 19 \\
267 & 8.6077 & 1CE3FF1EE6303981FF8EFDAD4D7A68754CC025E7A0C0F8AEDAA959B5366E92AB6C9 & 15 \\
269 & 8.5533 & 00044082B5CD9B863C953DA87B413F38D48FC9353AF1681D3F8D26919CDF028AEEAA & 15 \\
271 & 8.5496 & 03E68F8220F8E3C441F95788998DA34F71922C378D99DDA01A9444B6DAD775ADE6B5 & 13 \\
273 & 8.8221 & 00F0B85C0C72596C846774F322D11A054022AFEA1BBC2334F766E8C79E36CADE90B4A & 15 \\
275 & 8.5607 & 010183C7020637F8E56E0CC10AF4279B64CC639A742FD14CD6E06DAA3657524B59515 & 17 \\
277 & 8.6058 & 01F06C2C91741C9B351770FEC09E48014D889CFAA8E58AC54B77BF318DAF7B8C2C6B5A & 15 \\
279 & 8.5898 & 01DFE083C100C62740C4BA7F371E6249A1EE979C7669232A7BC4D42764D514B5D6A895 & 17 \\
281 & 8.5763 & 0756A543AB7350A4E93891673AA8A3ACB69CD870C120801367B89384E0BF37012FE07F6 & 17 \\
283 & 8.2925 & 000031BF9CC3A578A735BB90ED29E49F8E4EC6DA9C69F0ED1BB8327DA07B4C9AB935555 & 17 \\
285 & 8.5789 & 00C91658B0F1264C581A103A2FE15F6CAFCC6CD40C75FA54212BA1A9ECE63B4B09E7B8CA & 15 \\
287 & 8.3184 & 1E6B3C946F28C7790A9C847B7B2B02A302AFAFF537F53F3A3A45C9FD1A24DF2E41CB3E69 & 17 \\
289 & 8.1436 & 01907C03C37F71DF96C241D101BE8C97048EE448EB78C851ABBDAE2C795DB7572D2AD6B9A & 15 \\
291 & 8.1912 & 00FBF8380E3F809C03137D67359013846696CE1E645B15183260A313549D5AB6D5B5ABAD5 & 19 \\
293 & 8.0898 & 007843C21087843C09FFC1373B28EB311D6668667DBB304831373AD558AD2E968BA2D2E96A & 17 \\
295 & 8.0534 & 0780067FF006078033803C307E73B739CD9685E18C9B23B26A534B55B355A56552AA6555A5 & 17 \\
297 & 8.0190 & 07145E287D1AAAA8FA14F48F29B836694ECCC32CCC4F8672918348F4FA1480001BD6825EFB6 & 17 \\
299 & 8.1496 & 01FFEF631BD2188039CCC87C22C3FACF4E9A6BE79EC2CFAB4F74A5CCC9B55D970B9362EAA95 & 15 \\
301 & 8.3304 & 01F27C3817FF00910EDFB6F8C386C64E77537DDD73F764E6C692C94715C4BB8AB557A92D635A & 17 \\
303 & 8.0718 & 14A29CEB1AEA5CA5A43D99B1250A3155FEF3433432EA80137D0713998B4787C87EF93EC9F7C1 & 21 \\
\hline
\end{tabular}
\end{small}
}
\end{table*}

Table~\ref{tab:best:L201} presents skew-symmetric sequences of lengths 201 up to 303. 
For each length $L$, a sequence is presented using a hexadecimal notation. We decode each hexadecimal digit in binary form (0 $\mapsto$ 0000, 1 $\mapsto$ 0001, 2 $\mapsto$ 0010, \dots, F $\mapsto$ 1111), and, if necessary, remove initial
0 symbols to obtain a binary string of the appropriate length. Then we convert each 0 to $+1$ and each 1
to $-1$ to obtain the binary sequence. 
Table~\ref{tab:best:L201} contains binary sequences with merit factor $F>8$ and some of them have $F>9$. In this table, we show also the results up to $L=225$ that are taken from our previous work~\cite{Brest2018heuristic}, except for seven sequences with odd lengths of 205, 209, 215--221, where the merit factors have been improved by the impxLast algorithm.
For all other sequences longer or equal than 227 we report 
the merit factor values that are obtained by the impxLast and these results are currently the best known.

The largest sequence with a merit factor higher than 9 found to date has a length of 255. Interestingly, it could be pointed out that all sequences with $L<=285$, except for $L=283$, have the merit factor higher than 8.5.
One can notice that currently known best merit factors decrease as lengths of sequences become larger, and this decreasing trend can be seen by results obtained by our algorithm, but also by results of the lssOrel algorithm, as well as by Knauer's results. The reason for this lies in the fact that the search space is increasing exponentially. 
 
The authors believe that binary sequences with even higher merit factors exist, but one needs even more computational power to find them. Another possibility for searching new sequences with the higher merit factors is to invent new algorithms.\\
In Table~\ref{tab:best:L201}, a PSL value for each skew-symmetric binary sequence is shown in the last column. These PSL values are worse compared to the best-known PSL values (see~\cite{dimitrov2021hybrid,Brest2018heuristic}) for sequences with lengths $201\le L \le 303$. This means that a binary sequence with a high merit factor has a PSL value that is higher (i.e. worse) than the best-known PSL value, and vice versa. Consequently, a designer of a new algorithm should take care of this fact.

\section{Conclusion}
\label{sec:Conclusion}

In this paper, we used a stochastic algorithm and a high-performance computation to search for aperiodic binary sequences with low autocorrelation properties. We reported skew-symmetric binary sequences of length from 201 up to 303 that have the merit factor $F$ greater than 8, and many of them have $F>8.5$ and also $F>9$. The longest sequence with $F>9$ found to date has length $L=255$. In the future, our research will focus on searching new longer sequences with higher merit factors using parallel computation on  graphical processing units (GPUs). An implementation of a search algorithm using quantum operations~\cite{vzufan2021advances-Quantum} is also a possible direction of further research.


%
\begingroup
\makeatletter
\renewcommand\section{\@startsection {section}{1}{\z@}%
                                   {-3.5ex \@plus -1ex \@minus -.2ex}%
                                   {2ex \@plus.2ex}%
                                   {\large\sffamily\bfseries\color{mycolor}}}
\makeatother

\balance


\endgroup

\end{document}